\documentclass[aps,prl,preprint,superscriptaddress]{revtex4}
\usepackage{graphicx}
\usepackage{gensymb}
\begin{document}


\title{Linear polarization study of the microwave radiation-induced magnetoresistance-oscillations: comparison of power dependence to theory}


\author{Tianyu Ye}
\affiliation{Department of Physics and Astronomy, Georgia State University, Atlanta, Georgia 30303, USA}
\author{Jes\'us I\~narrea}
\affiliation{Escuela Polit\'ecnica Superior, Universidad Carlos III, Leganes, Madrid, Spain,}
\affiliation{Unidad Asociada al Instituto de Ciencia de Materiales, CSIC,
Cantoblanco,Madrid,28049,Spain.}
\author{W. Wegscheider}
\affiliation{Laboratorium f\"ur Festk\"orperphysik, ETH Z\"urich, 8093 Z\"urich, Switzerland}
\author{R. G. Mani}
\affiliation{Department of Physics and Astronomy, Georgia State University, Atlanta, Georgia 30303, USA}

\date{\today}

\begin{abstract}
We present an experimental study of the microwave power and the linear polarization angle dependence of the
microwave induced magnetoresistance oscillations in the high-mobility GaAs/AlGaAs two-dimensional electron system. Experimental
results show sinusoidal dependence of the oscillatory magnetoresistance extrema as a function of the polarization angle. Yet, as the microwave power increases, the angular dependence includes additional harmonic content, and begins to resemble the absolute value of the cosine
function. We present a theory to explain such peculiar behavior.
\end{abstract}


\maketitle

\section{introduction}
Microwave radiation-induced zero-resistance states\cite{Maninature2002, ZudovPRLDissipationless2003} are an interesting phenomenon in two dimensional electron systems (2DES) because, for example, coincidence of microwave-induced zero-resistance states with quantum Hall zero-resistance states leads to the extinction of the associated quantum Hall plateaus.\cite{ManiPRBPhaseStudy2009} The microwave radiation-induced magneto-resistance oscillations that lead into the zero-resistance states include characteristic traits such as periodicity in $B^{-1}$,\cite{Maninature2002,ZudovPRLDissipationless2003} a 1/4-cycle shift,\cite{ManiPRLPhaseshift2004}, distinctive sensitivity to temperature and microwave power,\cite{ManiPRBAmplitude2010,InarreaPRBPower2010} along with other specific features.\cite{KovalevSolidSCommNod2004,ManiPRBVI2004,SimovicPRBDensity2005,ManiPRBTilteB2005,WiedmannPRBInterference2008,ArunaPRBeHeating2011,ManiPRBPolarization2011,Fedorych2010,Wiedmann2011,Dai2011,RamanayakaPolarization2012,ManinatureComm2012,TYe2013,Inarrea2014reemission,Mani2013sizematter,ManiNegRes2013,TYe2014combine,ArunaPhysicaB,ManiPRBterahertz2013,TYe2014APL,HCLiu2014,TYe2015SciRep,Kvon2013,Chepelianskii2014,Chakraborty2014,Levin2015,Chepelianskii2015}. The linear-polarization-sensitivity of these oscillations has been a topic of intensive recent experimental study \cite{ManiPRBPolarization2011,RamanayakaPolarization2012,Lei2012Polar,Inarrea2013Polar,TYe2014combine,TYe2014APL,TYe2015SciRep}; these studies have shown that the amplitude of microwave radiation-induced magnetoresistance oscillations changes periodically with the linear microwave polarization angle.

Looking at the linear polarization characteristics in greater detail, at fixed temperature and polarization angle the amplitude of microwave radiation-induced magnetoresistance oscillations increases with the microwave power. It follows, approximately,\cite{ManiPRBAmplitude2010,InarreaPRBPower2010} $A=A_0P^{\alpha}$, where $A$ is the amplitude of microwave radiation-induced magnetoresistance oscillations, $A_0$ and $\alpha \approx 1/2$ are constants and $P$ is microwave power. At fixed temperature and microwave power, amplitude of microwave radiation-induced magnetoresistance oscillations changes sinusoidally with the linear polarization angle. The experimental results have shown that the longitudinal resistance $R_{xx}$ vs linear polarization angle $\theta$ follows a cosine square function\cite{RamanayakaPolarization2012}, i.e., $R_{xx}(\theta)=A \pm Ccos^2(\theta-\theta_0)$ ($A$ and $C$ are constants, $\theta_0$ is the phase shift, which depend on microwave frequency\cite{HCLiu2014}), at low microwave power. Note tha this angular dependence can also be rewritten as $R_{xx}(\theta)=A \pm (C/2)(1+cos[2(\theta-\theta_0)]$. Deviation from this functional form was noted for higher microwave intensities. 

From the theoretical point of view\cite{DurstPRLDisplacement2003,AndreevPRLZeroDC2003,RyzhiiJPCMNonlinear2003,KoulakovPRBNonpara2003,LeiPRLBalanceF2003,InarreaPRLeJump2005,DmitrievPRBMIMO2005,LeiPRBAbsorption+heating2005,ChepelianskiiEPJB2007,Inarrea2008,ChepelianskiiPRBedgetrans2009,Inarrea2011,Inarrea2012,Kunold2013,Zhirov2013,Lei2014Bicromatic,Yar2015, Ibarra-Sierra2015, Raichev2015}, there are many approaches to understand the physics of the microwave radiation-induced magnetoresistance oscillations. These include the radiation-assisted indirect inter-Landau-level scattering by phonons and impurities (the displacement model)\cite{DurstPRLDisplacement2003,LeiPRLBalanceF2003}, the periodic motion of the electron orbit centers under irradiation (the radiation driven electron orbit model)\cite{InarreaPRLeJump2005}, and a radiation-induced steady state non-equilibrium distribution (the inelastic model)\cite{DmitrievPRBMIMO2005}. Among these approaches, the radiation driven electron orbit model intensively considered temperature\cite{Inarrea2010Temp,Inarrea2005temp}, microwave power\cite{InarreaPRBPower2010} and microwave polarization direction\cite{Inarrea2013Polar} as factors that could change the amplitude of microwave radiation-induced magnetoresistance oscillations. Here, we report an experimental study of microwave power and linear polarization angle dependence of the radiation-induced magnetoresistance oscillations and compare the results with the predictions of the radiation driven electron orbit model. The comparison provides new understanding of the experimental results at high microwave powers.

\section{theoretical model}
The radiation driven electron orbit model\cite{InarreaPRLeJump2005,Inarrea2005temp,Inarrea2015holes,Inarrea2006} was developed to explain
the observed diagonal resistance, $R_{xx}$, of an irradiated 2DES at low magnetic field, $B$, with $B$ in the z-direction, perpendicular
to the 2DES. Here, electrons behave as 2D quantum oscillator in the $XY$ plane that contains the 2DES. The
system is also subjected to a DC electric field in the x-direction ($E_{DC}$), 
the transport direction, and microwave radiation that is linearly polarized at different angles ($\theta$)
with respect to the transport direction ($x$-direction). The radiation electric field  is given by
$\overrightarrow{E}(t)=(E_{0x} \overrightarrow{i}+E_{0y}
\overrightarrow{j})\cos wt$
where $E_{0x}$, $E_{0y}$ are the amplitudes  of the MW field and $w$
the frequency. Thus, $\theta$ is given by
$\tan \theta= \frac{E_{0y}}{E_{0x}}$.
The corresponding electronic hamiltonian can be exactly solved\cite{InarreaPRLeJump2005,Inarrea2007polar}
obtaining a solution for the total wave function,
\begin{equation}
\Psi(x,y,t)\propto\phi_{N}\left[(x-X-a(t)),(y-b(t)),t\right]
\end{equation}
where $\phi_{N}$ are Fock-Darwin states,  $X$ is the center of the orbit for the electron
motion, and
$a(t)$ (for the x-coordinate) and $b(t)$ (for the y-coordinate) are
the solutions for a $classical$ driven 2D harmonic oscillator
(classical uniform circular motion). The expressions for
 an arbitrary angle $\theta$ are given by
\begin{eqnarray}
a(t)&=&\left[\frac{\sqrt{w^{2}\cos^{2}
\theta + w_{c}^{2}\sin^{2} \theta}}{w}\right]\frac{eE_{0}\cos
wt}{m^{*}\sqrt{(w_{c}^{2}-w^{2})^{2}+\gamma^{4}} }=A_{x}\cos wt  \nonumber\\
\\
b(t)&=&\left[\frac{w\sqrt{w^{2}\cos^{2}
\theta + w_{c}^{2}\sin^{2} \theta}+(w_{c}^{2}-w^{2})\cos\theta}{ww_{c}}\right]\frac{eE_{0}\sin
wt}{m^{*}\sqrt{(w_{c}^{2}-w^{2})^{2}+\gamma^{4}} }=A_{y}\sin wt
\end{eqnarray}
where $e$ is the electron charge, $\gamma$ is a
 damping factor for the electronic
interaction with  acoustic phonons, $w_{c}$ the cyclotron
frequency and $E_{0}$ is the total amplitude of the radiation electric field.
These two latter equations give us the equation of an ellipse, $\frac{a^{2}}{A_{x}^{2}}+\frac{b^{2}}{A_{y}^{2}}=1$.
 Then, the first finding of
this theoretical model is that according to the expressions for $a(t)$ and $b(t)$, the center of the electron orbit
performs a classical
elliptical trajectory in the $XY$ plane driven by radiation  (see inset of Fig. 4).
This is  reflected in the
x and y directions  as harmonic oscillatory motions with the same frequency
as radiation.
In this elliptical motion electrons in their orbits interact with the lattice ions being damped and
emitting acoustic phonons; in the $a(t)$ and $b(t)$ expressions, $\gamma$
represents this damping.

The above expressions for $a(t)$ and $b(t)$ are obtained for a infinite 2DES.
But if we are dealing with finite samples the expressions can be slightly different because
the edges can play an important role.
Then the key issue of symmetry/asymmetry of the sample has to be considered  in regards of the polarization
sensitivity. In the experiments the
samples were rectangular-shaped or Hall bars (asymmetric samples)  and the $R_{xx}$ measurements were obtained at each  of the longest sides of the
sample between  two lateral contacts. According to this experimental set up, we observe that along the
classical elliptical trajectories the driven motion in the $x$ direction presents
fewer restrictions since the top and bottom sample edges are far from the
$R_{xx}$ measurement points (side contacts). Thus, there is no $spatial$ constraints  due to the existence of edges.
 However for the driven motion in the $y$ direction, the restrictions
are very important due to the presence of the edges from the very first moment. The lateral edges impede
or make more difficult the
MW-driven classical motion of the electron orbits in the $y$ direction.
The effect is as if the  $E_{0y}$ component of the microwave electric field
were much less efficient in coupling and driving the electrons than the $E_{0x}$ component.
This situation has to be reflected in the $a(t)$ and $b(t)$ expressions.
Thus, we have phenomenologically introduced an asymmetry factor $\lambda$ to deal
with this important scenario affecting the obtained $R_{xx}$.
Since for a more intense radiation electric field $E_{0}$, the motion in $y$ is increasingly hindered
we have introduced $\lambda= \frac{1}{1+c E_{0}}$ where $c$ is a constant that tends to $0$ for
symmetric samples and then $a(t)$ reads:
\begin{equation}
a(t)=\left[\sqrt{\cos^{2}\theta + \frac{w_{c}^{2}}{w^{2}} \left( \frac{1}{1+c E_{0}}\right)  \sin^{2} \theta}\right]\frac{eE_{0}
}{m^{*}\sqrt{(w_{c}^{2}-w^{2})^{2}+\gamma^{4}} }\cos
wt=A^{*}_{x} \cos
wt
\end{equation}


This $radiation-driven$ behavior has a deep impact on the charged
impurity scattering and in turn in the conductivity. Thus, first we
calculate the impurity scattering rate $W_{N,M}$
between two driven Landau states $\Psi_{N}$, and
$\Psi_{M}$\cite{InarreaPRLeJump2005,Inarrea2015holes}.
In order to calculate the electron drift velocity, next  we find the average effective distance advanced by the electron
in every scattering jump\cite{Inarrea2015holes}: $\Delta X^{MW}= \Delta X^{0}-A^{*}_{x}\sin w\tau$,
where $\tau$ is
the flight time that is strictly the time it takes the electron
to go from the initial orbit to the final one. This time
is part of the scattering time, $\tau_{S}$,  that is normally defined as
the average time between scattering events and  equal to
the inverse of the scattering rate. $\Delta X^{0}$ is the average advanced distance
without radiation. Finally the longitudinal conductivity
$\sigma_{xx}$ is given by:
$\sigma_{xx}\propto \int dE \frac{\Delta X^{MW}}{\tau_{S}}$
being $E$
the energy.
To obtain $R_{xx}$ we use
the relation
$R_{xx}=\frac{\sigma_{xx}}{\sigma_{xx}^{2}+\sigma_{xy}^{2}}
\simeq\frac{\sigma_{xx}}{\sigma_{xy}^{2}}$, where
$\sigma_{xy}\simeq\frac{n_{i}e}{B}$ and $\sigma_{xx}\ll\sigma_{xy}$.
Therefore,
\begin{equation}
R_{xx}\propto - \left[\sqrt{\cos^{2}\theta + \frac{w_{c}^{2}}{w^{2}} \left( \frac{1}{1+c E_{0}}\right)  \sin^{2} \theta}\right]\frac{eE_{0}
}{m^{*}\sqrt{(w_{c}^{2}-w^{2})^{2}+\gamma^{4}} }   \sin w\tau
\end{equation}

\section{experiments and results}
\begin{figure}[t]
\centering
\includegraphics[width= 150mm]{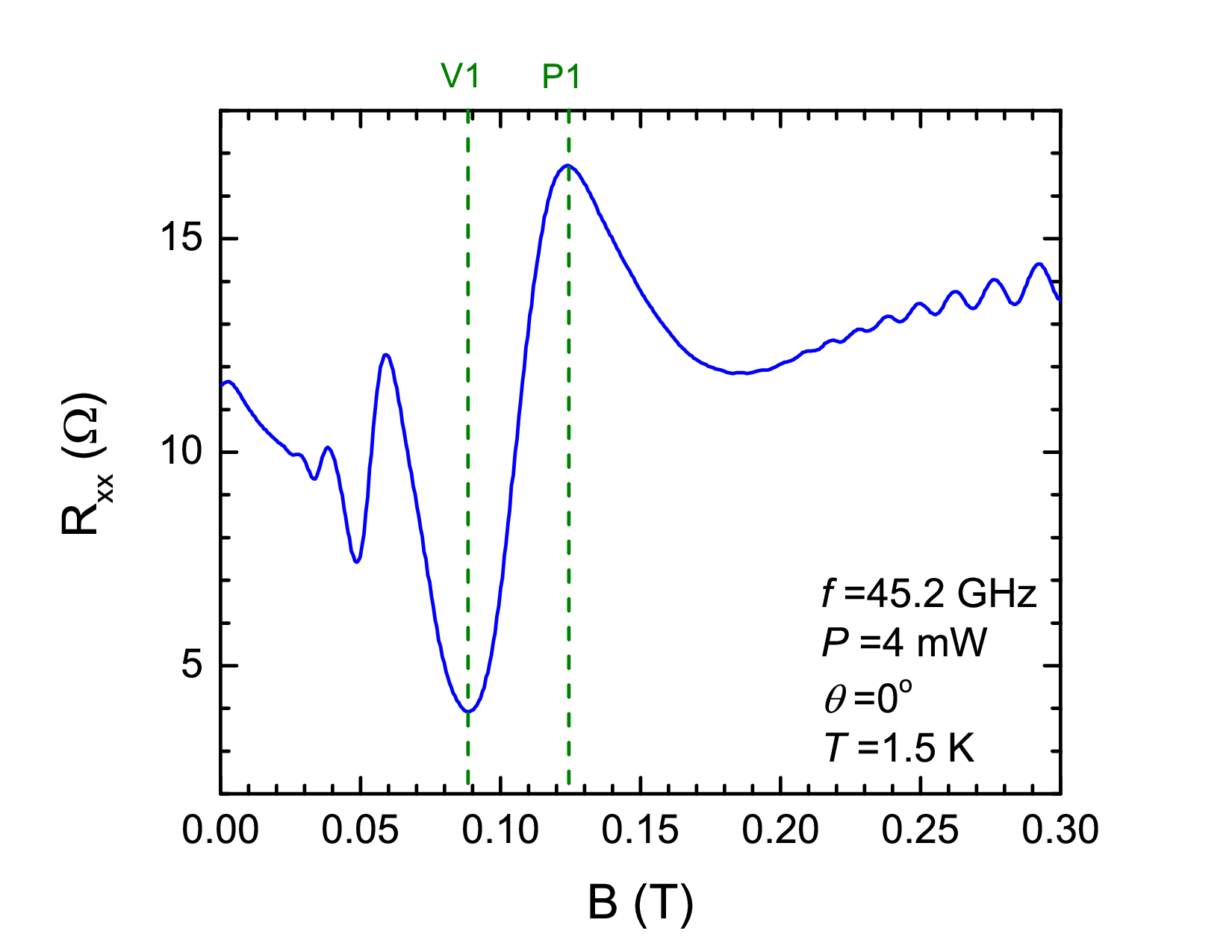}
\caption{(Color online) Longitudinal resistance $R_{xx}$ versus magnetic field $B$ with microwave photo-excitation at 45.2 GHz, 4 mW and $T$=1.5 K. The polarization angle, $\theta$, is zero. The labels, $P1$ and $V1$ at the top abscissa mark the magnetic
fields of the first peak and valley of the oscillatory magneto-resistance.}
\end{figure}

Experiments were carried out on high mobility GaAs/AlGaAs hetero-structure Hall bar samples. The samples were placed on a long cylindrical waveguide sample holder and loaded into a variable temperature insert (VTI) inside the bore of a superconducting solenoid magnet.  The high mobility condition was achieved in the 2DES by brief illumination with a red light-emitting diode at low temperature. A microwave launcher at the top of the sample holder excited microwaves within the cylindrical waveguide. The angle between the long axis of Hall bar sample and the antenna in  the microwave launcher is defined as the linear polarization angle. This linear polarization angle could be changed by rotating the microwave launcher outside the cryostat. Low frequency lock-in techniques were utilized to measure the diagonal and off-diagonal response of the sample. 

At 1.5 K, the longitudinal resistance $R_{xx}$ vs magnetic field $B$, exhibits strong microwave radiation-induced magnetoresistance oscillations, see Fig. 1, for frequency $f$ = 45.2 GHz, source power, $P$ =4 mW, and vanishing linear polarization, i.e. $\theta = 0$.  The figure shows that maxima and minima up to the fourth order are observable below 0.15 T. The first maxima and minima are designated as $P1$ and $V1$. In figs. 2 and 3, we examine the linear polarization angle dependence and the microwave power dependence at these extrema.

\begin{figure}[t]
\centering
\includegraphics[width= 120mm]{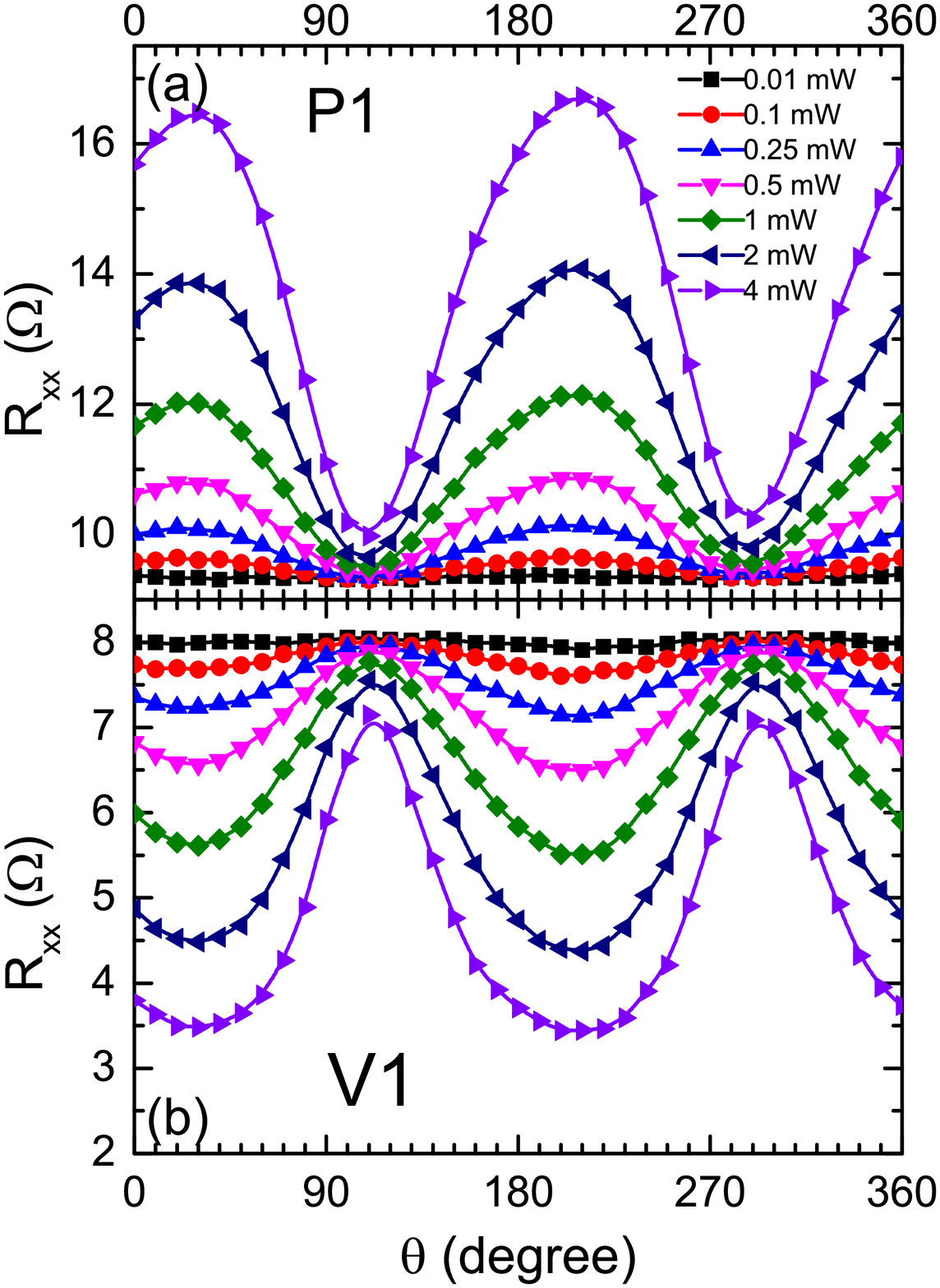}
\caption{(Color online) Longitudinal resistance $R_{xx}$ versus linear polarization angle $\theta$ at the magnetic field corresponding to (a) $P1$ and (b) $V1$. The microwave frequency is 42.5 GHz. Different colored symbols represent different source microwave powers from 0 to 4 mW.
}
\end{figure}

Figure 2 exhibits $R_{xx}$ vs linear polarization angle $\theta$ at different microwave powers at $P1$ and $V1$. The common features of the data at different microwave powers  are: a) they exhibit an oscillatory lineshape and b) the peaks and valleys at all powers occur at the same angle. At low microwave power, the oscillating curve could be represented by a simple sinusoidal function. However, as microwave power increases, the amplitude of the oscillatory curves increases and, at the same time,  deviations from the sinusoidal profile become observable and more prominent. For instance, at $P$= 4 mW, the maxima are relatively rounded and minima are relatively sharp for $P1$ and, in contrast, the maxima are sharp and minima are rounded for $V1$. At the other oscillatory extrema in the $R_{xx}$ vs $B$ trace at lower $B$, see Fig. 1, such deviations from the simple sinusoidal behavior were more prominent at the same power.

\begin{figure}[t]
\centering
\includegraphics[width= 120mm]{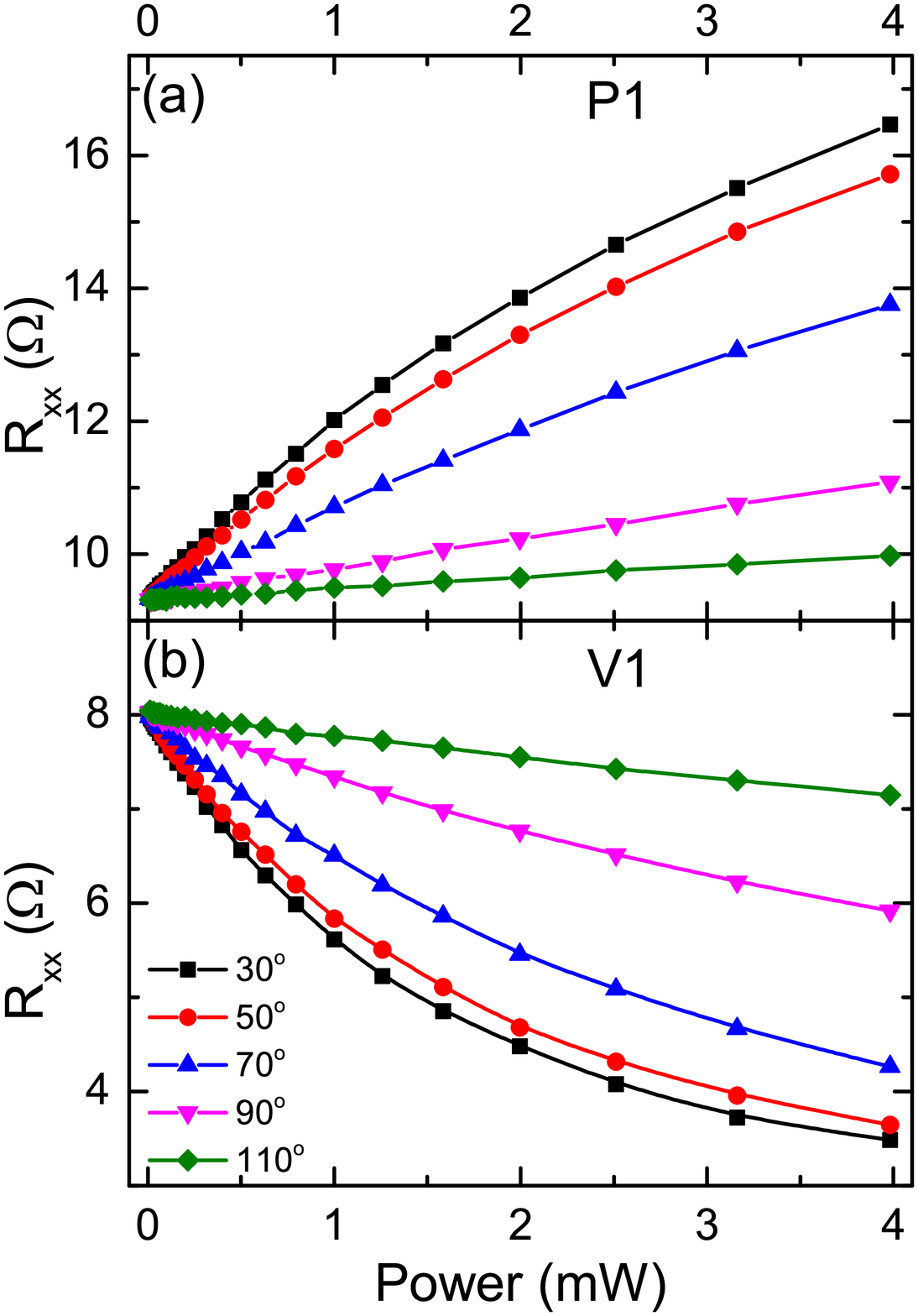}
\caption{(Color online) Figure exhibits the longitudinal resistance $R_{xx}$ versus microwave power $P$ at the magnetic field corresponding to (a) $P1$ and (b) $V1$. The microwave frequency is $f$ = 42.5 GHz. Different color symbols represent different linear polarization angles, $\theta$, between $30^{0}$ and $110^{0}$.}
\end{figure}

Figure 3 shows $R_{xx}$ vs. $P$ at different linear polarization angle $\theta$ for $P1$ and $V1$. For $P1$, see Fig. 3(a), $R_{xx}$ increases non-linearly as the microwave power increases. On the other hand, see Fig. 3(b), $R_{xx}$  decreases non-linearly with increasing $P$ at $V1$. In Fig. 3(a) and (b), all traces start at the same resistance value at $P$=0.01 mW since the (essentially) dark resistance is invariant under polarization angle rotation. 

\section{calculated results}
\begin{figure}[t]
\centering
\includegraphics[width= 120mm]{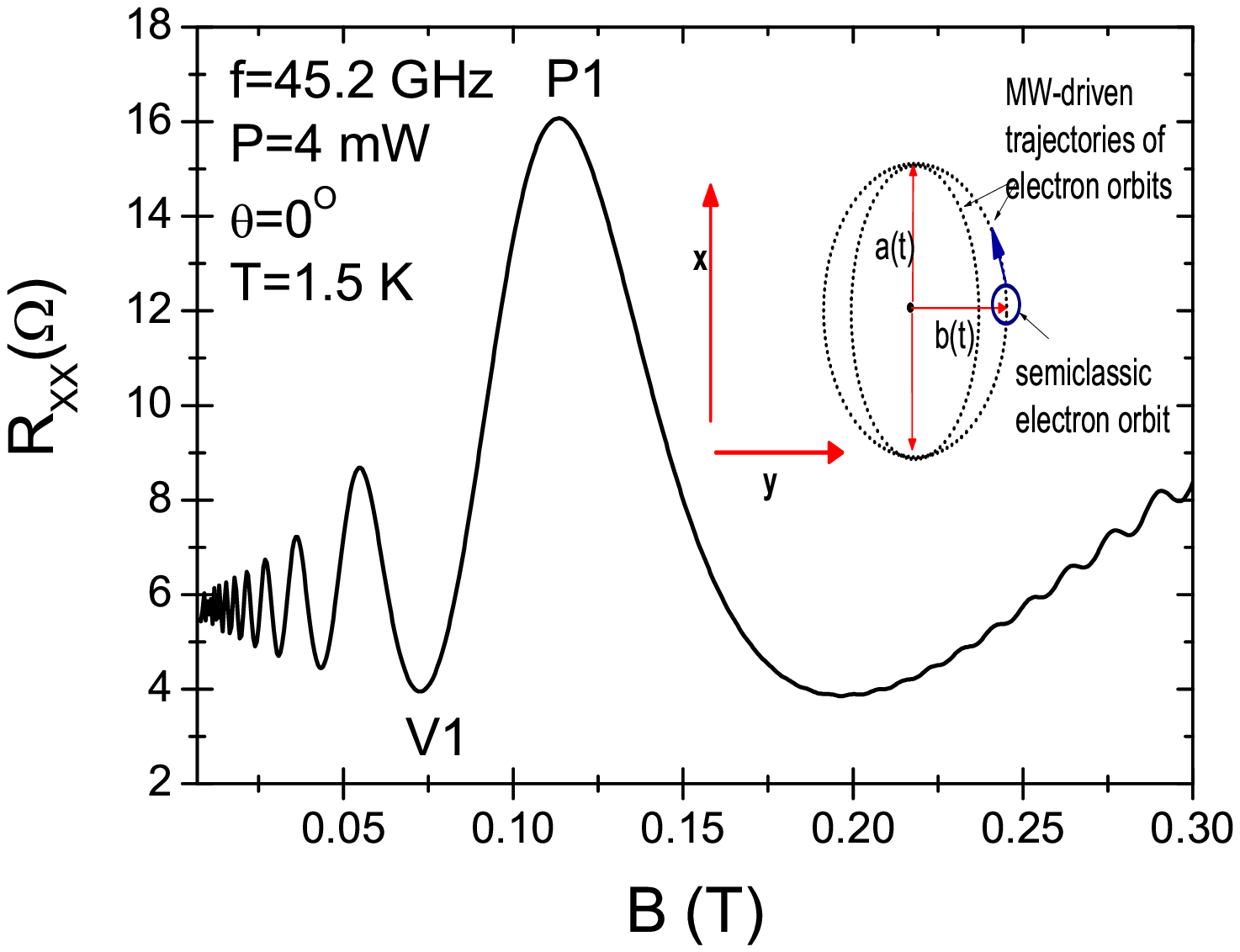}
\caption{(Color online) Calculated irradiated magnetoresistance $R_{xx}$ versus magnetic field $B$ with microwave frequency of 45.2 GHz, microwave
power $P=4$ mW and $T$=1.5 K. The polarization angle, $\theta$, is zero. Symbols $P1$ and $V1$ correspond to the peak and valley, respectively, as indicated.}
\end{figure}
In Figure 4, we present calculated results of irradiated $R_{xx}$ versus $B$ for a microwave frequency of $45.2$ GHz and power of $P=4 mW$. As in the
experimental curve of Fig. 2, we obtain,  at a  temperature of $T=1.5K$,
  clear $R_{xx}$ oscillations which turn out to be qualitatively and quantitatively similar to experiment. For the exhibited curve, the polarization angle is zero. The most prominent peak and valley are labelled as
$P1$ and $V1$ respectively. In the inset of this figure we present  a schematic diagram showing the radiation-driven classical trajectories
of the guiding center of the electron orbit.

Figure 5 exhibits calculated results of irradiated $R_{xx}$  versus linear polarization angle $\theta$ for different microwave powers for  peak
$P1$ and valley $V1$ in panels a) and b) respectively. The microwave frequency is 45.2 GHz and $T$ is 1.5 K. The microwave power ranges from 0.01 to 6.0 mW.
As in the experimental results of Figure 2, we observe that the $R_{xx}$ curves evolve from a clear sinusoidal profile  at low microwave powers to
a different profile where, for instance, in upper panel ($P1$) the peaks broaden and
the valleys get sharpened. Similar trend is observable in the lower panel ($V1$). The explanation for
this peculiar behavior can be obtained from  equation [5] and the
square root between brackets. When the microwave power (electric field amplitude $E_{0}$) increases,
the factor $\frac{w_{c}^{2}}{w^{2}} \left( \frac{1}{1+c E_{0}}\right)  \sin^{2} \theta$ gets smaller and smaller.
As  a result, the $R_{xx}$ curve begins to lose its simple sinusoidal profile. At high powers the latter factor
is so small that the $\cos^{2}$ factor is predominant and the square root tends to
the absolute value of $\cos^{2}$:
\begin{equation}
 \left[\sqrt{\cos^{2}\theta + \frac{w_{c}^{2}}{w^{2}} \left( \frac{1}{1+c E_{0}}\right)  \sin^{2} \theta}\right]
 \rightarrow \sqrt{\cos^{2}\theta} = abs[\cos \theta ]
\end{equation}
The profile observed in experiment and calculation for high powers is very similar
to the one of the $abs[\cos \theta ]$. This evolution can be clearly observed in the
panel (c) of Figure 5.

In Figure 6, we present calculated $R_{xx}$ under radiation versus the microwave power for
different polarization angles for peak $P1$, upper panel, and for valley $V1$, lower panel.
For increasing angles from $0^{0}$ to $90^{0}$ the behavior of $P1$ and $V1$ are similar
in the sense that the corresponding intensity of both becomes smaller and smaller. In other
words, for increasing angles the height of the peak decreases and the depth of the valley
decreases too. The theoretical explanation comes from the square root term as before.
For increasing angles the cosine term tends to zero becoming predominant the sine term.
However the latter gets smaller for increasing power due to the asymmetry factor.
All curves share one important feature, the non-linearity of $R_{xx}$
versus $P$ suggesting a sublinear relation. This  behavior can be
straightforward explained with our model in terms of:
\begin{equation}
 E_{0}\propto \sqrt{P}\Rightarrow  R_{xx}\propto \sqrt{P}.
\end{equation}
And then the  $R_{xx}$ profile with $P$ follows a square root dependence
as experiments show.


\begin{figure}[t]
\centering
\includegraphics[width= 80mm]{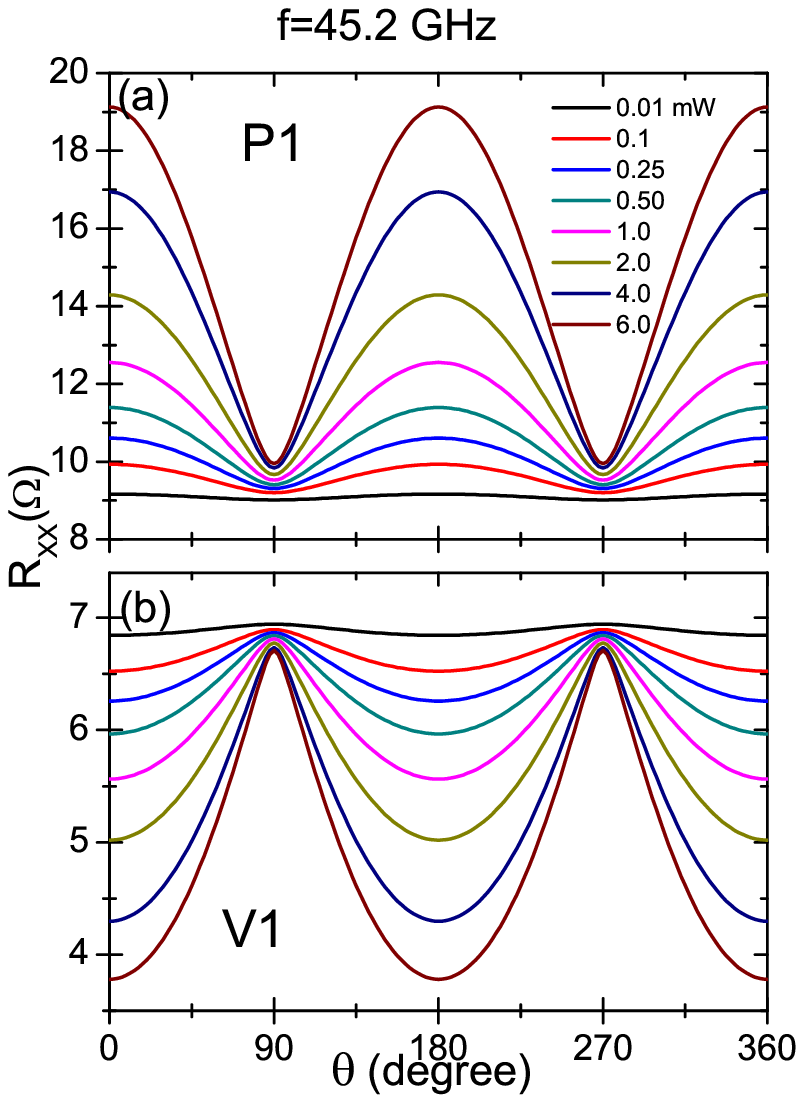}
\centering
\includegraphics[width=80mm]{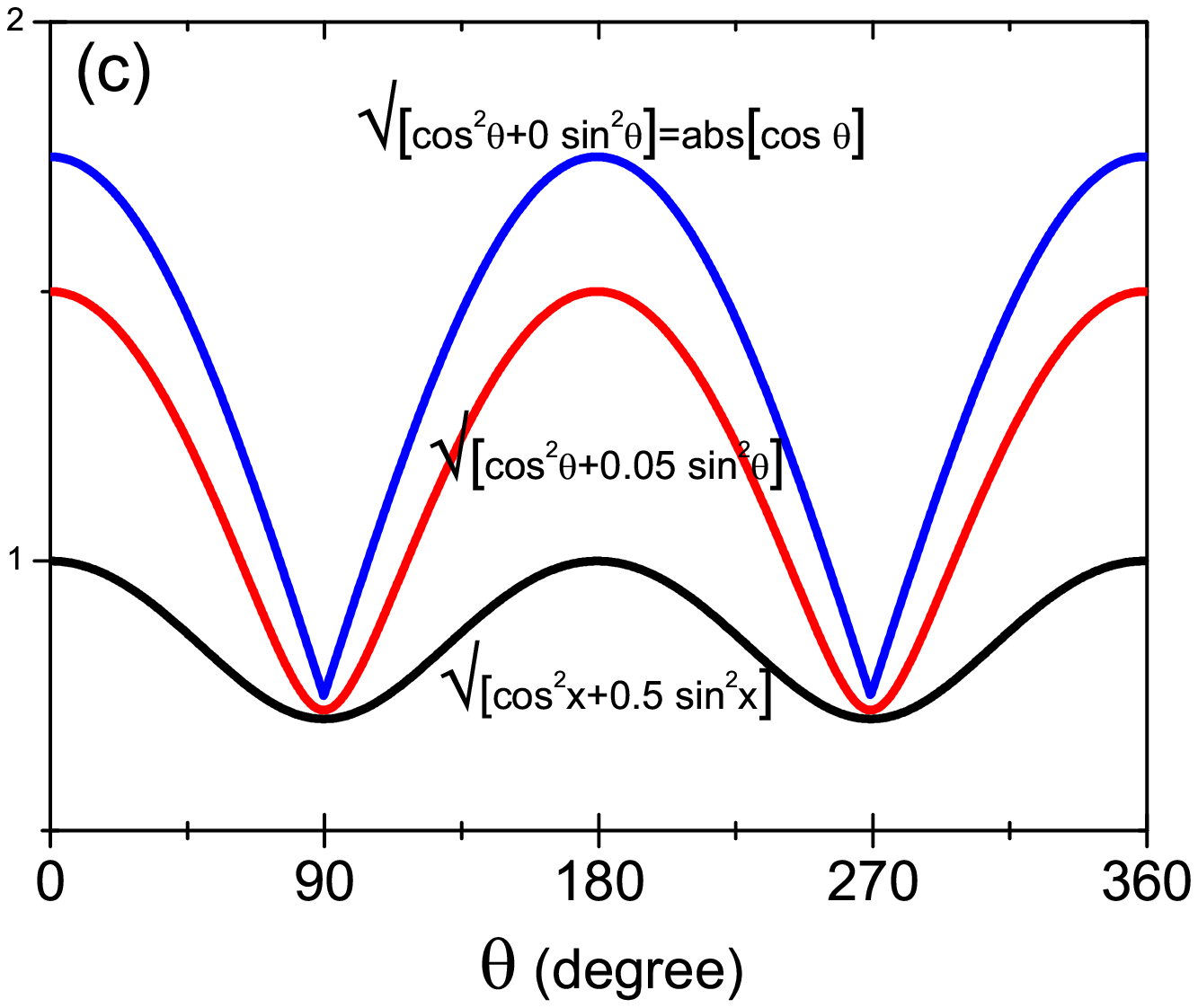}
\caption{(Color online) Calculated irradiated  magnetoresistance $R_{xx}$ versus linear polarization
 angle $\theta$ for the magnetic fields corresponding to the peak $P1$ (panel (a)) and valley $V1$ (panel (b))
 In panel (c) we present simulation of the curves evolution with mathematical functions when the
 $\sin^{2}$ term decreases.
The microwave photo-excitation at 45.2 GHz, the microwave power ranges from from 0.01 to 6.0 mW and $T$=1.5 K. }
\end{figure}

\begin{figure}[t]
\centering
\includegraphics[width= 100mm]{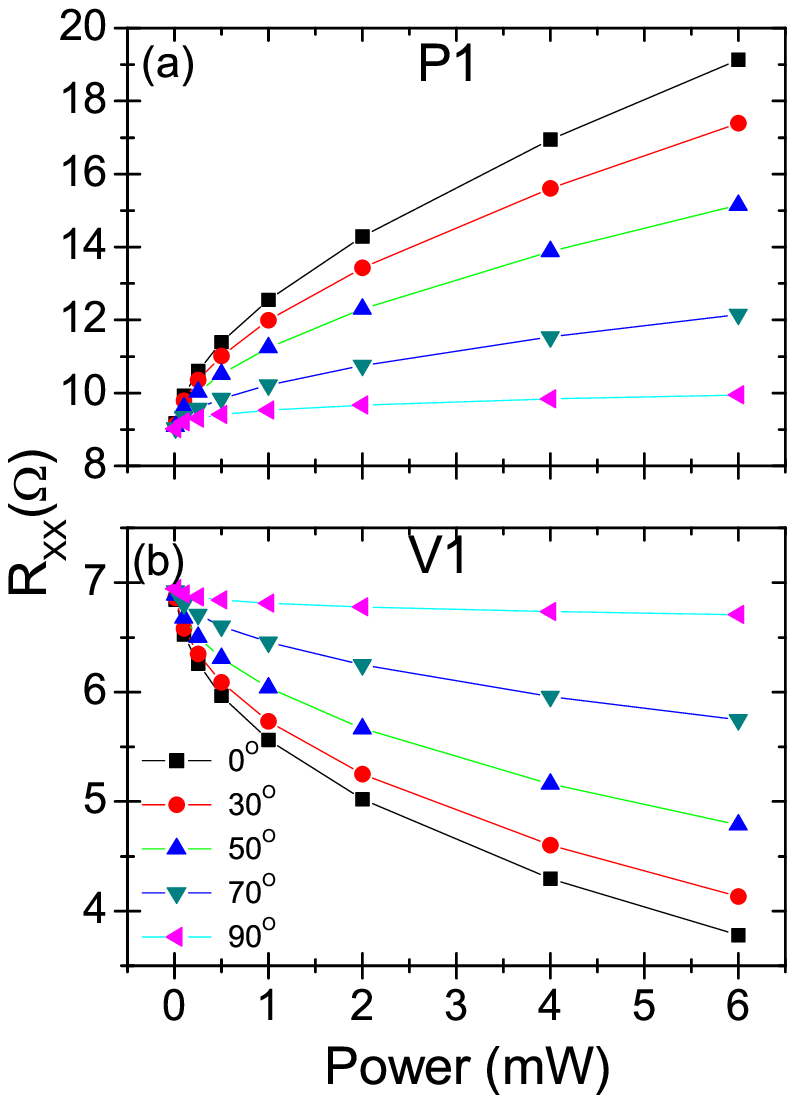}
\caption{(Color online) Calculated irradiated magnetoresistance  $R_{xx}$ versus microwave power.
The microwave frequency is 45.2 GHz and $T$=1.5 K. The polarization angle, $\theta$, ranges from $0^{0}$ to $90^{0}$, as indicated.}
\end{figure}

\section{conclusion}
We have presented  together experimental- and theoretically-calculated- results concerning the microwave power and linear polarization angle dependence of the
microwave irradiated oscillatory magnetoresistance in the GaAs/AlGaAs two-dimensional electron system. Experimental
results show that, as the microwave power increases, the $R_{xx}$ vs. $\theta$ traces, see Fig. 2, 
gradually lose the simple sinusoidal profile, as the profile begins to resemble the absolute value of the cosine function. We presented the theoretical insight to explain this evolution using the radiation-driven electron orbit model, which suggests a profile following $abs[\cos \theta ]$.

Intuitively, one can motivate the change in the profile of the $R_{xx}$ vs. $\theta$ curves with increasing power by noting that increasing the power, i.e., driving the 2DES more strongly, most likely increases the harmonic content in the $R_{xx}$ vs. $\theta$ lineshape, which leads to deviations from the simple sinusoidal function proposed for low power in ref. \cite{RamanayakaPolarization2012}. In the low power limit, this theoretical prediction of the radiation driven electron orbit model matched the experimental suggestion of $R_{xx}(\theta)=A \pm Ccos^2(\theta-\theta_0) = A \pm (C/2)(1+cos[2(\theta-\theta_0)]$ \cite{RamanayakaPolarization2012}. 

One might tie together the low power  $R_{xx}$ vs. $\theta$ lineshape ($\sim cos^2(\theta)$)  with the high power $R_{xx}$ vs. $\theta$ lineshape ($abs[\cos \theta ] $) by examining the Fourier expansion over the interval $[-\pi, \pi]$ of $abs[\cos \theta ] = (2/\pi) + (4/\pi) \sum_{m=1}^{\infty} (((-1)^{m}/(1-4m^{2}))cos(2m\theta)$. If one keeps the lowest order ($m=1$) term, which is the only one likely to be observable at low power, then $abs[\cos \theta ] \rightarrow (2/\pi)(1  + (2/3)cos(2\theta) ) \sim cos^2(\theta)$.

\section{acknowledgement}
Magnetotransport measurements at Georgia State University are
supported by the U.S. Department of Energy, Office of Basic Energy
Sciences, Material Sciences and Engineering Division under
DE-SC0001762. Additional support is provided by the ARO under
W911NF-07-01-015. J.I.  is supported by the MINECO (Spain) under grant
MAT2014-58241-P  and ITN Grant 234970 (EU).
GRUPO DE MATEMATICAS APLICADAS A LA MATERIA CONDENSADA, (UC3M),
Unidad Asociada al CSIC.

\pagebreak

\section*{Figure Captions}
Figure 1)  Longitudinal resistance $R_{xx}$ versus magnetic field $B$ with microwave photo-excitation at 45.2 GHz, 4 mW and $T$=1.5 K. The polarization angle, $\theta$, is zero. The labels, $P1$ and $V1$ at the top abscissa mark the magnetic
fields of the first peak and valley of the oscillatory magneto-resistance.

Figure 2) Longitudinal resistance $R_{xx}$ versus linear polarization angle $\theta$ at the magnetic field corresponding to (a) $P1$ and (b) $V1$. The microwave frequency is 42.5 GHz. Different colored symbols represent different source microwave powers from 0 to 4 mW.

Figure 3) Figure exhibits the longitudinal resistance $R_{xx}$ versus microwave power $P$ at the magnetic field corresponding to (a) $P1$ and (b) $V1$. The microwave frequency is $f$ = 42.5 GHz. Different color symbols represent different linear polarization angles, $\theta$, between $30^{0}$ and $110^{0}$.

Figure 4) Calculated irradiated magnetoresistance $R_{xx}$ versus magnetic field $B$ with microwave frequency of 45.2 GHz, microwave
power $P=4$ mW and $T$=1.5 K. The polarization angle, $\theta$, is zero. Symbols $P1$ and $V1$ correspond to the peak and valley, respectively, as indicated.

Figure 5) Calculated irradiated  magnetoresistance $R_{xx}$ versus linear polarization
 angle $\theta$ for the magnetic fields corresponding to the peak $P1$ (panel (a)) and valley $V1$ (panel (b))
 In panel (c) we present simulation of the curves evolution with mathematical functions when the
 $\sin^{2}$ term decreases.
The microwave photo-excitation at 45.2 GHz, the microwave power ranges from from 0.01 to 6.0 mW and $T$=1.5 K.

Figure 6)   Calculated irradiated magnetoresistance  $R_{xx}$ versus microwave power.
The microwave frequency is 45.2 GHz and $T$=1.5 K. The polarization angle, $\theta$, ranges from $0^{0}$ to $90^{0}$, as indicated.
\pagebreak

\begin{figure}[t]
\centering
\includegraphics[width= 120mm]{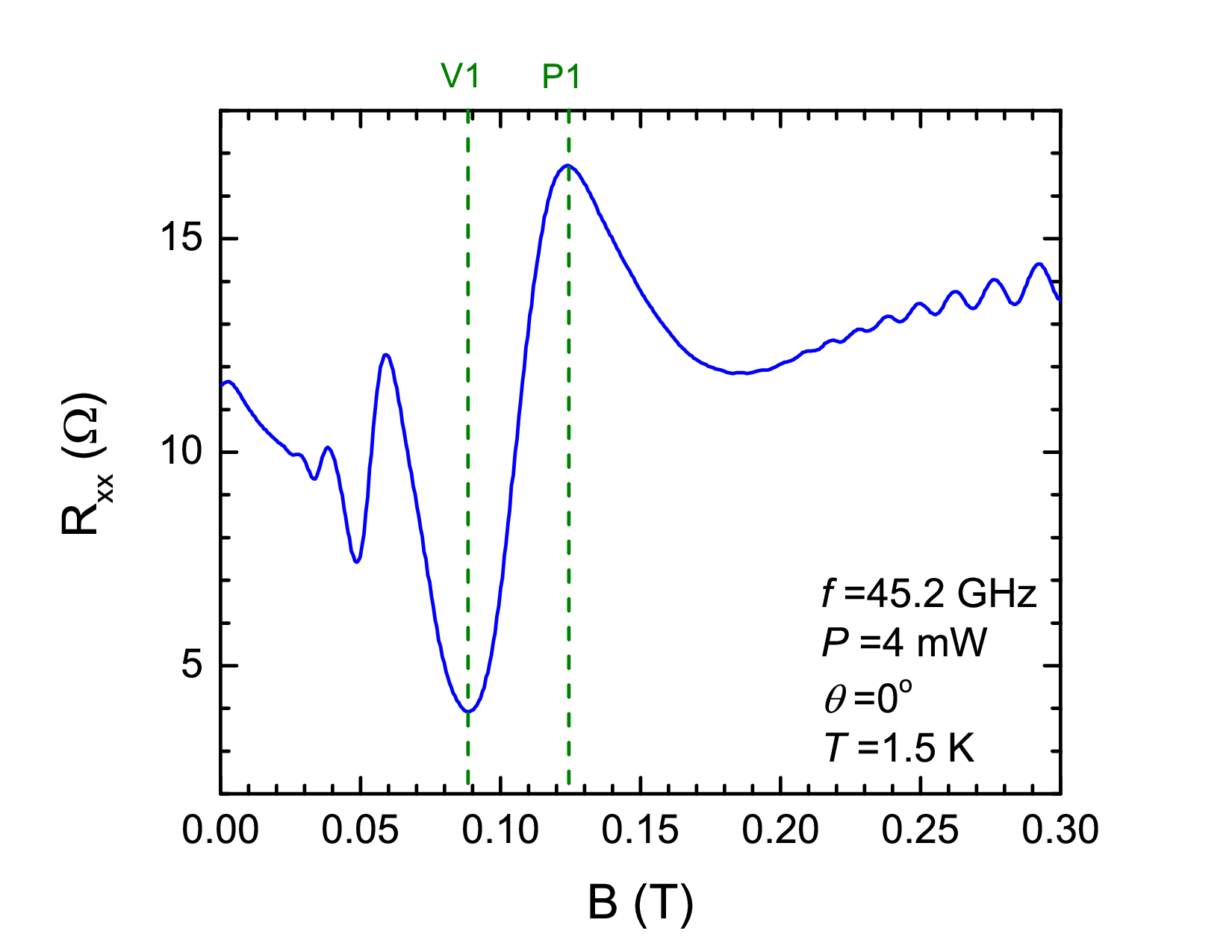}
\begin{center}
Figure 1
\end{center}
\end{figure}

\begin{figure}[t]
\centering
\includegraphics[width=120 mm]{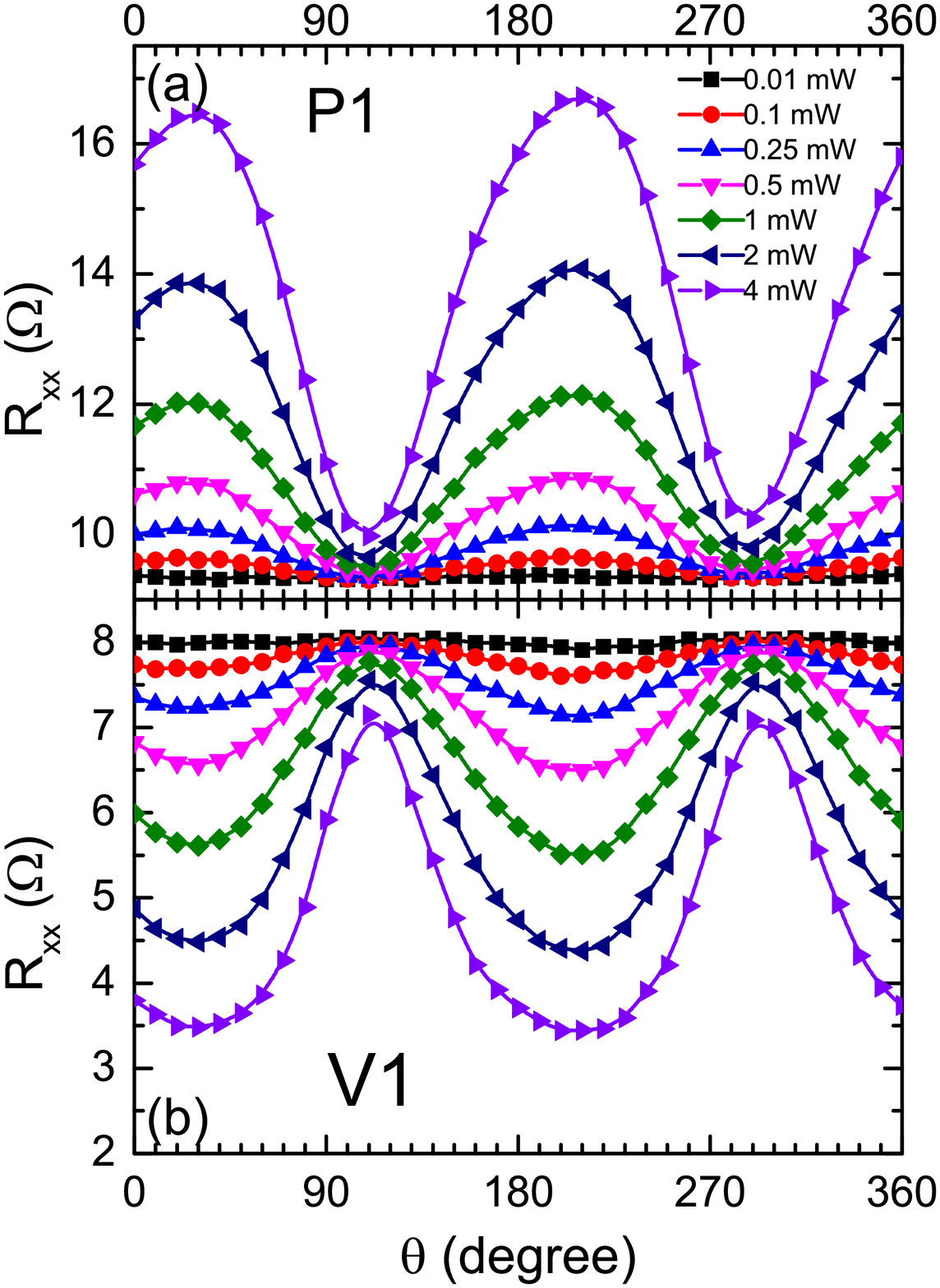}
\begin{center}
Figure 2
\end{center}
\end{figure}

\begin{figure}[t]
\centering
\includegraphics[width= 120mm]{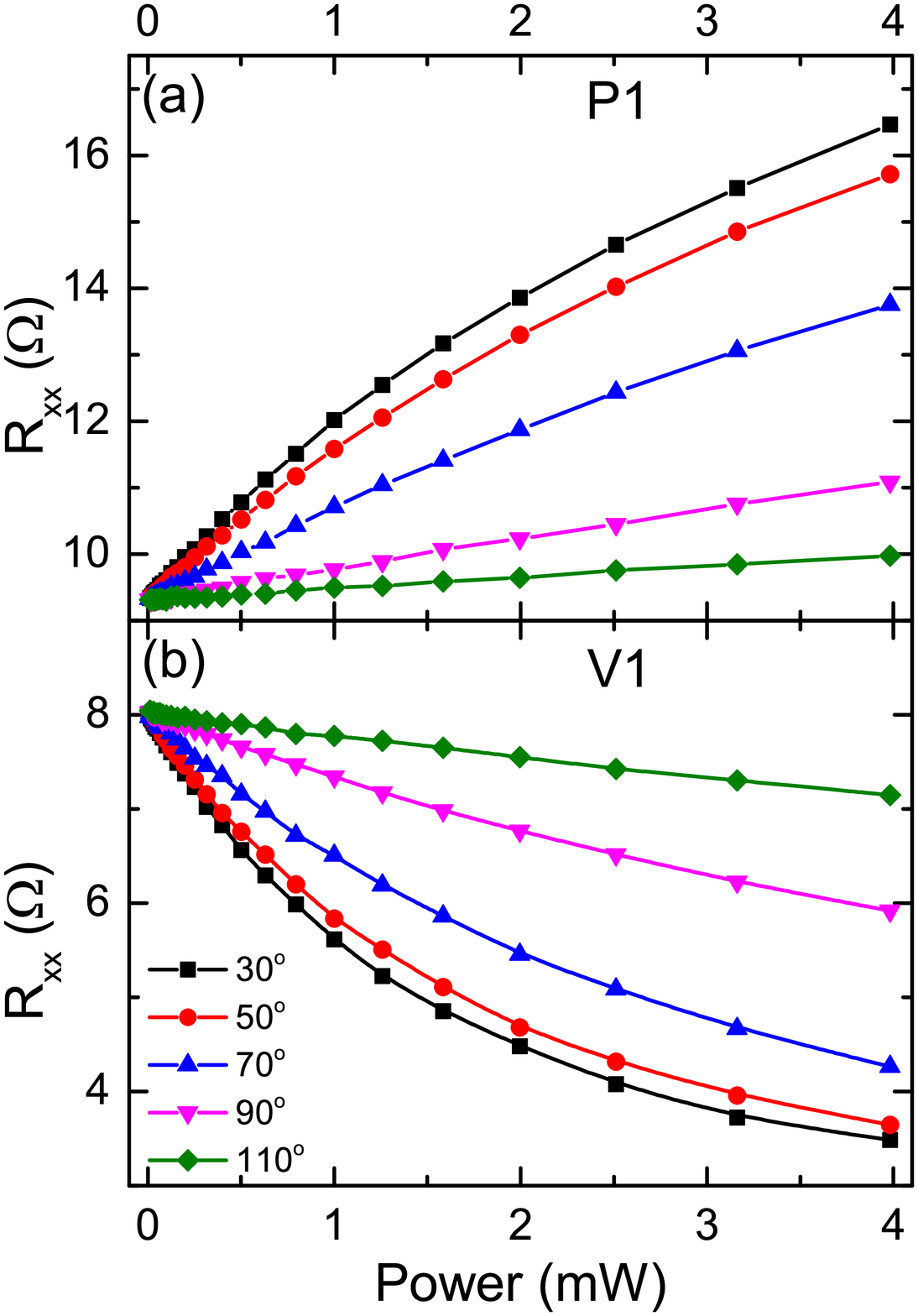}
\begin{center}
Figure 3
\end{center}
\end{figure}

\begin{figure}[t]
\centering
\includegraphics[width= 120mm]{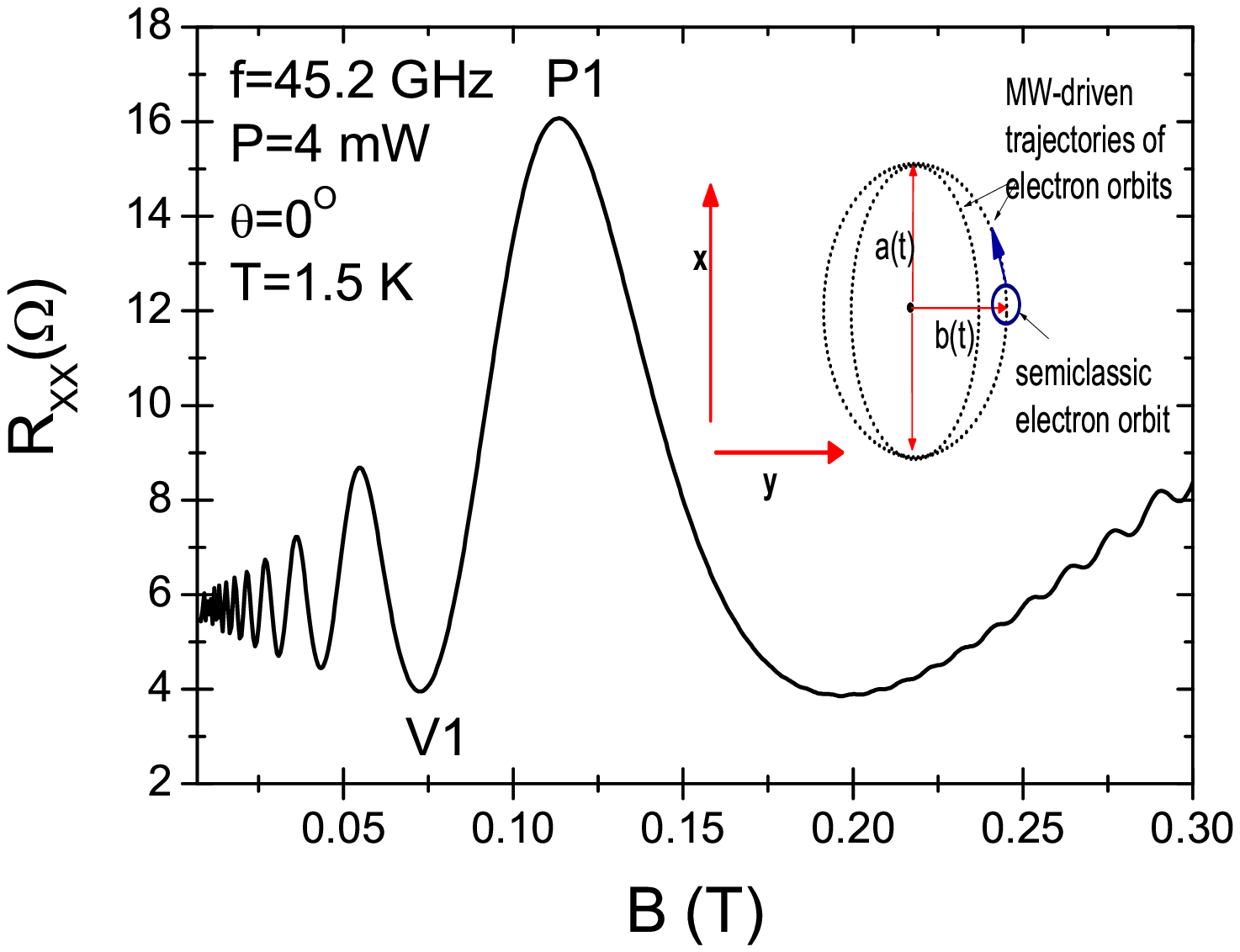}
\begin{center}
Figure 4
\end{center}
\end{figure}

\begin{figure}[t]
\centering
\includegraphics[width= 120mm]{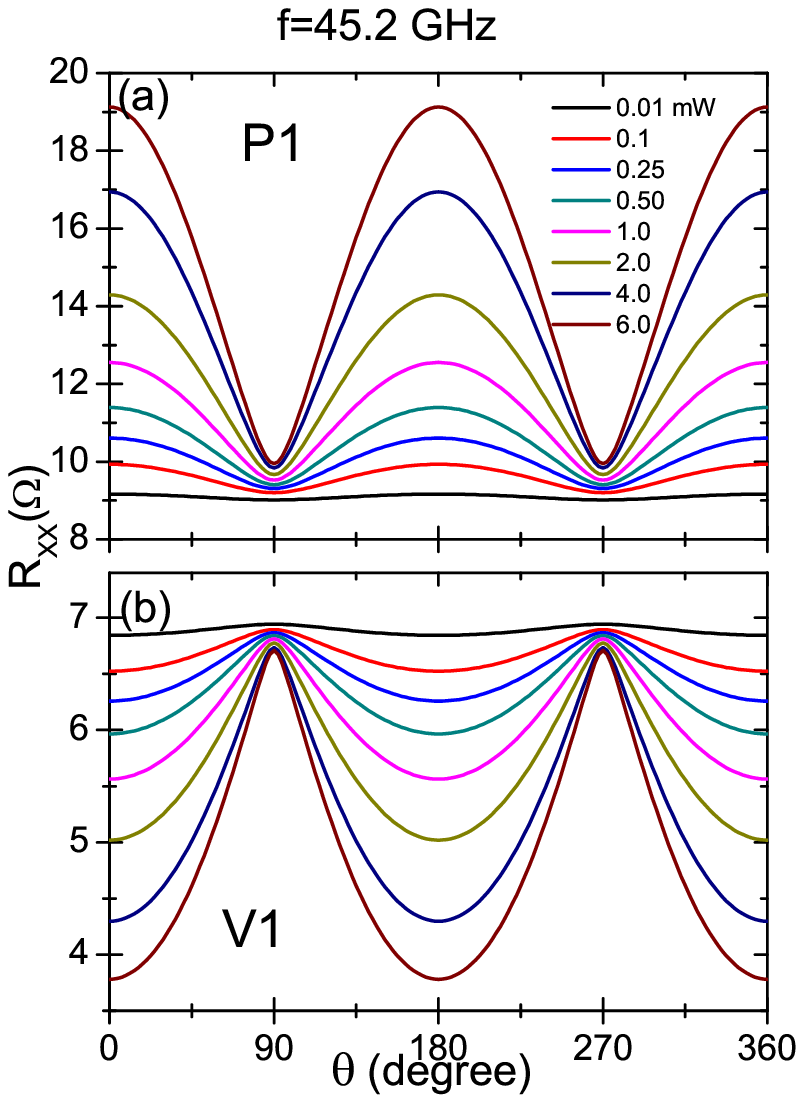}
\centering
\includegraphics[width=80mm]{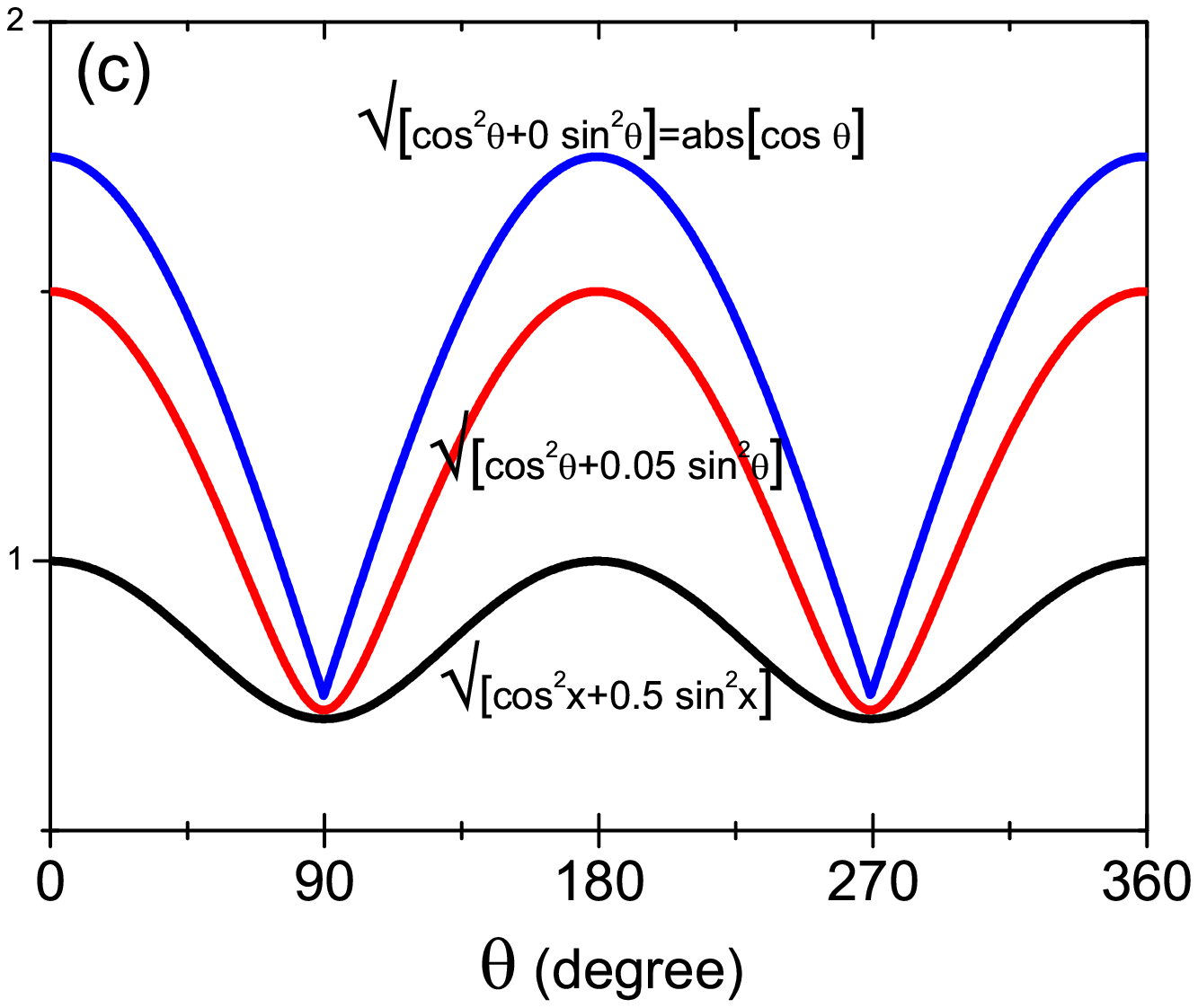}
\begin{center}
Figure 5
\end{center}
\end{figure}

\begin{figure}[t]
\centering
\includegraphics[width= 120mm]{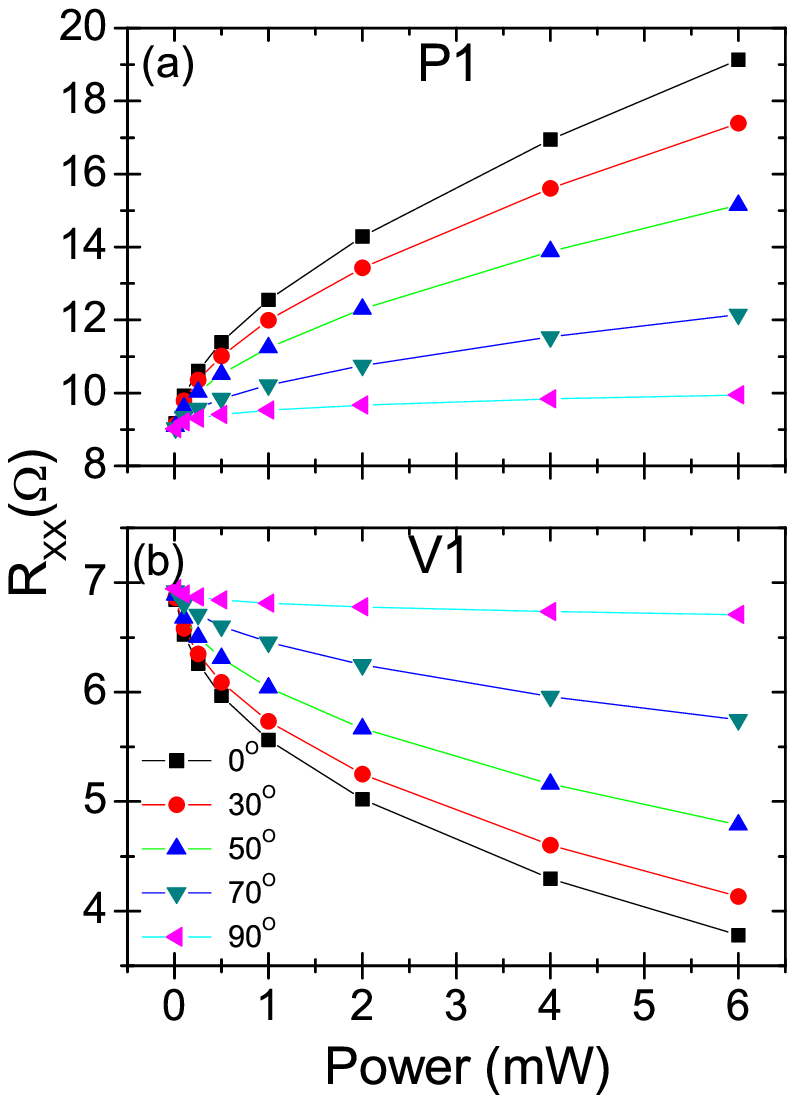}
\begin{center}
Figure 6
\end{center}
\end{figure}

\end{document}